\documentclass[conference]{IEEEtran}
\IEEEoverridecommandlockouts
\usepackage{cite}
\usepackage{amsmath,amssymb,amsfonts}
\usepackage{algorithm}
\usepackage{algpseudocode}
\usepackage{graphicx}
\usepackage{subfigure}
\usepackage{caption}
\usepackage{float}
\usepackage{textcomp}
\usepackage{xcolor}
\usepackage{booktabs}
\def\BibTeX{{\rm B\kern-.05em{\sc i\kern-.025em b}\kern-.08em
    T\kern-.1667em\lower.7ex\hbox{E}\kern-.125emX}}
\begin{document}

\title{ Meta-Learning Driven Lightweight Phase Shift Compression for IRS-Assisted Wireless Systems
}

\author{

\IEEEauthorblockN{Xianhua Yu\IEEEauthorrefmark{1}, Dong Li\IEEEauthorrefmark{2}, Bowen Gu\IEEEauthorrefmark{3}, Xiaoye Jing\IEEEauthorrefmark{1}, Wen Wu\IEEEauthorrefmark{4}, Tuo Wu\IEEEauthorrefmark{5} and Kan Yu\IEEEauthorrefmark{2}\IEEEauthorrefmark{6}}
\IEEEauthorblockA{\IEEEauthorrefmark{1} School of Electrical Engineering and Intelligentization, Dongguan University of Technology, Dongguan 523808, China}
\IEEEauthorblockA{\IEEEauthorrefmark{2} School of Computer Science and Engineering, Macau University of Science and Technology, Macau, China}
\IEEEauthorblockA{\IEEEauthorrefmark{3} School of Computer Science and Technology, Xinjiang University, Urumqi, 830046, China}
\IEEEauthorblockA{\IEEEauthorrefmark{4}  Frontier Research Center, Peng Cheng Laboratory, Shenzhen, 518055, China}
\IEEEauthorblockA{\IEEEauthorrefmark{5}  School of Electrical and Electronic Engineering, Nanyang
 Technological University, 639798, Singapore}
\IEEEauthorblockA{\IEEEauthorrefmark{6}  Key Laboratory of Universal Wireless Communications, Beijing University of Posts and Telecommunications, Beijing, China}
\IEEEauthorblockA{Email: xianhuacn@foxmail.com}
}

\maketitle

\begin{abstract}
The phase shift information (PSI) overhead poses a critical challenge to enabling real-time intelligent reflecting surface (IRS)-assisted wireless systems, particularly under dynamic and resource-constrained conditions. In this paper, we propose a lightweight PSI compression framework, termed meta-learning-driven compression and reconstruction network (MCRNet). By leveraging a few-shot adaptation strategy via model-agnostic meta-learning (MAML), MCRNet enables rapid generalization across diverse IRS configurations with minimal retraining overhead. Furthermore, a novel depthwise convolutional gating (DWCG) module is incorporated into the decoder to achieve adaptive local feature modulation with low computational cost, significantly improving decoding efficiency. Extensive simulations demonstrate that MCRNet achieves competitive normalized mean square error performance compared to  state-of-the-art baselines across various compression ratios, while substantially reducing model size and inference latency. These results validate the effectiveness of the proposed asymmetric architecture and highlight the practical scalability and real-time applicability of MCRNet for dynamic IRS-assisted wireless deployments.
\end{abstract}

% \begin{IEEEkeywords}
% Intelligent reflecting surface, phase shift compression, deep learning
% \end{IEEEkeywords}

\section{Introduction}
The evolution toward beyond-5G (B5G) and 6G networks is propelled by emerging applications such as extended reality (XR), autonomous driving, and massive IoT, which demand ultra-low latency, high reliability, and ubiquitous connectivity~\cite{XShen,BGu,KYu1}. Intelligent reflecting surfaces (IRS) have gained significant attention as a cost- and energy-efficient solution to meet these requirements by reconfiguring wireless propagation without additional active radio-frequency (RF) chains. An IRS comprises a large array of passive elements capable of adjusting the phase of incident electromagnetic waves, enabling precise control of the wireless environment. By intelligently tuning these phase shifts, IRS can enhance link quality, expand coverage, and improve spectral and energy efficiency~\cite{DLi1,DLi2,HXie1,HXie2,TWu,JHe}, all without increasing transmit power or hardware complexity.

To fully exploit the potential of IRS-assisted wireless systems, the phase shift information (PSI) must be reliably and efficiently transmitted from the base station (BS) to the IRS controller, ensuring accurate adjustment of each reflecting element. However, achieving accurate PSI delivery is often challenged by the control signaling overhead\cite{YChoi, XYi, XLin,XYu2}. In particular,  for IRS arrays comprising hundreds or even thousands of reflecting elements, transmitting the complete PSI as control signaling over a bandwidth-limited control channel imposes prohibitive signaling overhead. Moreover, this problem becomes even more severe in dynamic environments, where user mobility, environmental changes, or IRS reconfigurations necessitate frequent PSI updates. 

% Efficient compression and low-latency reconstruction of PSI are therefore crucial to enable practical, real-time deployment of IRS-assisted wireless systems.

To alleviate the PSI overhead bottleneck, recent studies\cite{XYu2,XYu1,ZLi,HFeng} have explored various deep learning-based compression methods utilizing autoencoders, aiming to learn compact latent representations by compressing the PSI at the BS side and reconstructing it at the IRS side. The proposed S-GAPSCN in \cite{XYu2} adopts an asymmetric structure, where the architecture of the decoder. In \cite{XYu1},  a denoising module is introduced in the PSCDN's decoder to alleviate the noise effect. An adaptive compression autoencoder-based model named ACFNet was proposed in \cite{ZLi}. A knowledge-based autoencoder, PSFNet, was proposed in \cite{HFeng} to further reduce the compression rate.

While these approaches achieve notable compression gains, they may suffer from two critical limitations: high computational complexity and poor generalization to unseen environments. Firstly, unlike the BS, which has sufficient computational resources, the IRS controller can only handle simple tasks, requiring low computational complexity in the decoder. Secondly, conventional deep learning models are typically task-specific and often require retraining when environmental conditions change, otherwise leading to significant performance degradation. However, training existing methods typically requires large-scale, task-specific datasets, which incurs substantial storage and computational overhead, making them infeasible for practical IRS-assisted wireless systems operating under varying conditions.
 
Motivated by the limitations of existing PSI compression solutions, this paper proposes a meta-learning-driven framework designed to address both the computational constraints and dynamic adaptation challenges in IRS-assisted wireless systems. The main contributions of this work are summarized as follows:
\begin{itemize}
    \item We design a lightweight and meta-adaptive autoencoder, termed the meta compression and reconstruction network (MCRNet), which enables efficient PSI compression under asymmetric computational conditions between the BS and the IRS controller.
    
    \item We propose a novel depth-wise convolutional gating (DWCG) module integrated into the decoder, enabling adaptive local feature modulation. By dynamically modulating feature importance based on local spatial context, DWCG emphasizes informative features while attenuating irrelevant ones, thereby improving reconstruction accuracy and maintaining low complexity.
    
    \item We develop a few-shot adaptation framework based on model-agnostic meta-learning (MAML), enabling rapid generalization to diverse IRS configurations with significantly reduced training sample requirements.
    
    \item We conduct comprehensive simulations demonstrating that MCRNet achieves superior NMSE performance, significantly smaller model size, and faster inference latency compared to state-of-the-art lightweight baselines.
\end{itemize}

\section{System Model and Problem Formulation}
\subsection{System Model}

\begin{figure}
    \centering
    \includegraphics[width=0.6\linewidth]{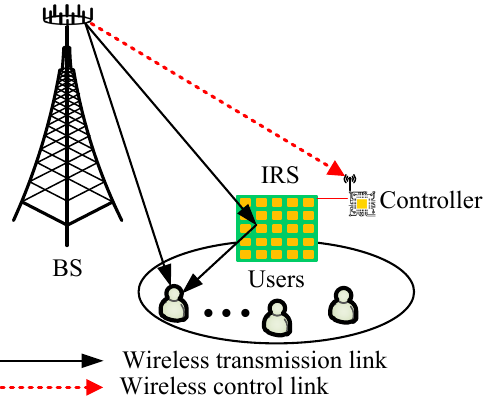}
    \caption{A typical downlink IRS-assisted multiple-input multiple-output (MIMO) wireless system}
    \label{IRS}
\end{figure}

We consider a downlink IRS-assisted multiple-input multiple-output (MIMO) wireless system, comprising a single BS, a single IRS, and \(K\) users, as illustrated in Fig.~\ref{IRS}. The BS is equipped with \(N_t\) transmit antennas, while each user is equipped with \(N_r\) receive antennas. The IRS consists of \(M\) passive reflecting elements, each capable of independently adjusting the phase shift of incident electromagnetic signals.
Let \(\mathbf{G} \in \mathbb{C}^{M \times N_t}\) denote the BS-to-IRS channel matrix, \(\mathbf{h}_{r,k} \in \mathbb{C}^{M \times 1}\) denote the IRS-to-user-\(k\) channel vector, and \(\mathbf{h}_{d,k} \in \mathbb{C}^{N_r \times N_t}\) denote the direct BS-to-user-\(k\) channel matrix. The BS transmits a signal \(\mathbf{s} \in \mathbb{C}^{N_t \times 1}\) with total transmit power \(P\). The received baseband signal at user \(k \in \{1, \dots, K\}\) is given by
\begin{equation}
    \mathbf{y}_k = \sqrt{P} \mathbf{h}_{r,k}^H \mathbf{\Phi} \mathbf{G} \mathbf{s} + \sqrt{P} \mathbf{h}_{d,k} \mathbf{s} + \mathbf{u}_k,
\end{equation}
where \(\mathbf{u}_k \sim \mathcal{CN}(\mathbf{0}, \sigma_k^2 \mathbf{I})\) denotes the additive white Gaussian noise. The IRS reflection matrix is modeled as
\begin{equation}
    \mathbf{\Phi} = \rho \cdot \mathrm{diag}\left(e^{j\theta_1}, e^{j\theta_2}, \dots, e^{j\theta_M}\right),
\end{equation}
where \(\rho \in (0,1]\) represents the reflection amplitude (typically close to unity), and \(\theta_m \in [0, 2\pi)\) denotes the phase shift induced by the \(m\)-th reflecting element.

Due to hardware constraints, the phase shifts \(\{\theta_m\}_{m=1}^M\) are uniformly quantized with \(K\) bits, resulting in the quantized set
\begin{equation}
    \mathcal{Q} = \left\{0, \frac{2\pi}{2^K}, \dots, \left(2^K-1\right)\frac{2\pi}{2^K}\right\}.
\end{equation}
The quantized phase shift vector is denoted by \(\mathbf{t} = [\theta_1, \theta_2, \dots, \theta_M]^T \in \mathcal{Q}^M\).

In this work, we aim to efficiently compress and reconstruct the phase shift vector \(\mathbf{t}\), thereby reducing the PSI transmission overhead and enabling practical deployment in resource-constrained IRS-assisted wireless systems.

\subsection{Problem Formulation}

To achieve low-overhead PSI delivery, we design an asymmetric autoencoder architecture, where the encoder \(f_{\text{enc}}(\cdot)\) is deployed at the BS with sufficient computational resources, and the decoder \(f_{\text{dec}}(\cdot)\) is implemented at the resource-constrained IRS controller.

Specifically, given the quantized phase vector \(\mathbf{t} \in \mathcal{Q}^M\), the encoder compresses it into a low-dimensional latent vector, which is transmitted over a control link possibly affected by noise and linear distortions. The decoder reconstructs the original phase vector from the received signal. The overall inference process is expressed as
\begin{equation}
    \hat{\mathbf{t}} = f_{\text{dec}}\left( \mathbf{H} \cdot f_{\text{enc}}(\mathbf{t}) + \mathbf{w} \right),
\end{equation}
where \(\mathbf{H}\) denotes a channel gain, and \(\mathbf{w} \sim \mathcal{N}(\mathbf{0}, \sigma_w^2 \mathbf{I})\) represents the additive Gaussian noise introduced during transmission.

The objective is to minimize the reconstruction error between the reconstructed phase vector \(\hat{\mathbf{t}}\) and the original \(\mathbf{t}\). The corresponding optimization problem is formulated as
\begin{equation}
    \min_{\Theta} \mathcal{L}(\Theta) = \mathbb{E} \left[ \left\| \mathbf{t} - f_{\Theta}(\mathbf{t}) \right\|^2 \right],
\end{equation}
where \(f_{\Theta}(\cdot) \triangleq f_{\text{dec}}\left( \mathbf{H} f_{\text{enc}}(\cdot) \right)\), and \(\Theta\) denotes the set of learnable parameters of the encoder and decoder.

\section{Proposed Framework}

This section presents our proposed framework for efficient PSI compression. The design consists of two key components: (i) a lightweight, meta-adaptive autoencoder named MCRNet, and (ii) a meta-learning-based few-shot adaptation mechanism for fast generalization across varying scenarios. We first introduce the MCRNet architecture designed for asymmetric deployment, followed by the MAML-based meta-training strategy.
\begin{figure*}
    \centering
    \includegraphics[width=0.95\linewidth]{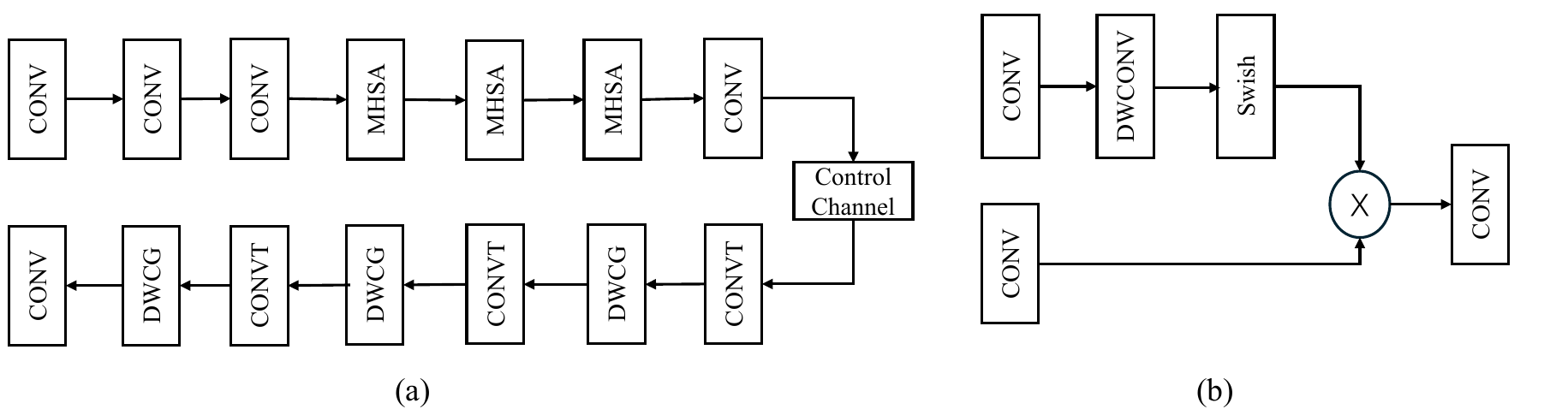}
    \caption{(a) The Architecture of MCRNet. The CONVT denotes the transpose convolution.(b) The Architecture of DWCG. The \(\otimes\) denotes element-wise multiplication}
    \label{framework}
\end{figure*}
\subsection{Preliminary Knowledge}

We briefly review the core building blocks adopted in MCRNet, including depthwise convolution, nonlinear activations, and multi-head self-attention mechanisms.

\subsubsection{Depthwise Convolution}

Depthwise convolution (DWConv) is a parameter-efficient operation that applies a distinct filter to each input channel, substantially reducing computational complexity compared to standard convolutions. Formally, the output feature map at spatial location \((i,j)\) for channel \(k\) is computed as
\begin{equation}
    y_{i,j,k} = \sum_m \sum_n w_{m,n,k} \cdot x_{i+m,j+n,k},
\end{equation}
where \(w_{m,n,k}\) denotes the kernel weight and \(x_{i+m,j+n,k}\) is the corresponding input feature. 

DWConv enables lightweight feature extraction and is particularly suitable for resource-constrained neural network architectures.

\subsubsection{Activation Functions: Swish and GELU}

Nonlinear activation functions are essential for enhancing the expressiveness of deep neural networks. The Swish activation function is defined as
\begin{equation}
    \text{Swish}(x) = x \cdot \sigma(\beta x),
\end{equation}
where \(\sigma(\cdot)\) denotes the sigmoid function and \(\beta\) is a learnable parameter. The sigmoid function is given by
\begin{equation}
    \sigma(x) = \frac{1}{1 + e^{-x}},
\end{equation}
which maps real-valued inputs to the range \([0,1]\).

Another commonly used activation function is the Gaussian Error Linear Unit (GELU), formulated as
\begin{equation}
    \text{GELU}(x) = x \cdot \Phi(x),
\end{equation}
where \(\Phi(x)\) represents the cumulative distribution function (CDF) of the standard Gaussian distribution.

Both Swish and GELU provide smoother nonlinearities compared to the widely used ReLU, and have demonstrated superior performance in lightweight deep network designs.

\subsubsection{Multi-Head Self-Attention (MHSA)}

MHSA mechanisms significantly enhance the model's capacity to capture global contextual dependencies by computing multiple attention distributions in parallel. Given input features, three key matrices are generated: query (\(Q\)), key (\(K\)), and value (\(V\)), through learned linear projections.

The overall MHSA operation is expressed as
\begin{equation}
    \text{MHSA}(Q,K,V) = \text{Concat}(\text{head}_1, \dots, \text{head}_h) W^O,
\end{equation}
where each attention head is computed as
\begin{equation}
    \text{head}_i = \text{Attention}(QW_i^Q, KW_i^K, VW_i^V),
\end{equation}
and \(W_i^Q\), \(W_i^K\), \(W_i^V\), and \(W^O\) are learnable projection matrices.

The scaled dot-product attention within each head is defined as
\begin{equation}
    \text{Attention}(Q, K, V) = \text{softmax}\left( \frac{QK^T}{\sqrt{d_k}} \right) V,
\end{equation}
where \(d_k\) denotes the dimensionality of the key vectors.

MHSA enables efficient modeling of long-range spatial interactions, which is crucial for preserving structural dependencies in PSI compression tasks.

\subsection{The Architecture of MCRNet}

To achieve efficient PSI compression under asymmetric hardware constraints, we propose a lightweight autoencoder-based architecture optimized for IRS-assisted wireless systems. The encoder, deployed at the BS side with sufficient computational resources, performs global feature extraction and semantic compression. In contrast, the decoder, operating at the resource-constrained IRS side, is designed for real-time, low-complexity reconstruction. This asymmetric design offloads the computational burden to the encoder, enabling lightweight, fast, and energy-efficient decoding at the IRS controller.

As shown in Fig.~\ref{framework}(a), the encoder consists of multiple convolutional layers and MHSA blocks, while the decoder consists of multiple transposed convolutional layers and DWCG modules\footnote{In this section, we take CR=$\frac{1}{8}$ as example.}.

The encoder takes as input a PSI matrix \(\mathbf{X} \in \mathbb{R}^{H \times W}\), which is first reshaped into a two-dimensional tensor \(\mathbf{X} \in \mathbb{R}^{HW \times 1}\). The reshaped tensor is successively processed by three convolutional layers with kernel size \(3\), stride \(2\), and padding \(1\). The first convolutional layer increases the channel dimension from \(1\) to \(C\), yielding a feature map \(\mathbf{F}_1 \in \mathbb{R}^{\frac{HW}{2} \times C}\). The second convolutional layer maintains the channel dimension while further reducing the spatial size to \(\mathbf{F}_2 \in \mathbb{R}^{\frac{HW}{4} \times C}\). The third convolutional layer further reduces the spatial size to \(\mathbf{F}_3 \in \mathbb{R}^{\frac{HW}{8} \times C}\). Each convolutional layer is followed by a GELU activation function to introduce nonlinearity.

To capture non-local spatial dependencies in the compressed representation, \(3\) stacked MHSA blocks are applied to \(\mathbf{F}_3\). The MHSA mechanism models long-range interactions across the feature map, enhancing the encoder’s ability to preserve structural and contextual information critical for accurate PSI reconstruction.

At the end of the encoder, a fully connected layer reduces the channel dimension to \(1\). The output of the encoder is a globally contextualized latent tensor \(\mathbf{z} \in \mathbb{R}^{\frac{HW}{8} \times 1}\), which is transmitted to the decoder side for reconstruction.

The decoder reconstructs the original PSI matrix from the latent representation through a combination of transposed convolutional upsampling and adaptive local feature modulation via the proposed Depthwise Convolutional Gating (DWCG) modules. It can be divided into two stages: an upsampling stage and a feature modulation stage.

During the upsampling stage, three sequential transposed convolutional layers are employed to progressively recover the spatial resolution. Each transposed convolution adopts a kernel size of 3, a stride of 2, padding of 1, and output padding of 1. Specifically, the first transposed convolution upsamples the latent tensor to \( z_1 \in \mathbb{R}^{\frac{HW}{4} \times C} \), the second to \( z_2 \in \mathbb{R}^{\frac{HW}{2} \times C} \), and the third to \( z_3 \in \mathbb{R}^{HW \times C} \). A ReLU activation function is applied after each transposed convolution to introduce nonlinearity and stabilize the upsampling process.

Following the upsampling stage, we perform adaptive local feature modulation via the proposed DWCG modules. Feature modulation refers to the process of dynamically adjusting the strength of feature representations based on local context, allowing the network to selectively emphasize informative patterns and suppress irrelevant ones at a fine-grained spatial level. Unlike self-attention mechanisms that globally model pairwise dependencies to assign importance weights across all spatial positions, DWCG focuses on local adaptivity with low computation cost, making it highly suitable for resource-constrained scenarios.

In the DWCG module, the input feature tensor is processed through a dual-branch structure: an activation branch and a value branch. Given an input vector \( z_{\text{in}} \in \mathbb{R}^{HW \times C} \), the activation branch applies a depthwise one-dimensional convolution with a kernel size of 3, followed by a Swish activation function, to generate the gating coefficients \( G \):
\begin{equation}
 G = \text{Swish}(\text{DWConv}_{k=3}(W_A * z_{\text{in}}))   
\end{equation}

Simultaneously, the value branch applies a depthwise one-dimensional convolution with a kernel size of 1 to obtain the modulated feature values \( V \):
\begin{equation}
    V = \text{DWConv}_{k=1}(W_V * z_{\text{in}})
\end{equation}

The final output \( \hat{z} \) is obtained by element-wise multiplication between the gating coefficients and the value features:
\begin{equation}
    \hat{z} = G \otimes V
\end{equation}

where \( \otimes \) denotes element-wise multiplication.

Finally, a \(1 \times 1\) convolution is applied to project the refined features to a single-channel output:
\begin{equation}
    \hat{X} = \text{Conv1}(\hat{z}) \in \mathbb{R}^{H \times W}
\end{equation}

Through this design, the decoder first restores the spatial structure and then adaptively modulates local features, ensuring accurate PSI reconstruction with extremely low computational overhead. By leveraging transposed convolutions for efficient resolution recovery and DWCG modules for lightweight adaptive feature modulation, the proposed decoder architecture effectively balances reconstruction accuracy, model compactness, and inference speed, thus meeting the stringent real-time requirements of IRS-assisted wireless systems.

\subsection{Meta-Learning Driven PSI Compression Scheme}

To achieve robust and efficient PSI compression across dynamic IRS-assisted wireless scenarios, we propose a meta-learning-based training framework that enables fast adaptation of a neural compression model to new environments using only a few support samples. We leverage the MAML approach to optimize the model across different conditions.

\subsubsection{Meta-Training Setup}

During meta-training, each task \(T_i\) corresponds to a specific IRS scenario, characterized by factors such as user distribution, channel realization, IRS element size, or phase quantization resolution. 

For each task, the dataset is split into:
\begin{itemize}
    \item \textbf{Support Set} \(D_i^{\text{support}}\): used for inner-loop adaptation.
    \item \textbf{Query Set} \(D_i^{\text{query}}\): used for outer-loop meta-update evaluation.
\end{itemize}

The support set typically contains \(K_s\) samples and the query set contains \(K_q\) samples, randomly sampled from the task data.

\subsubsection{Meta-Optimization Formulation}

The meta-learning objective is to find a model initialization \(\theta\) that can rapidly adapt to any new task after a few gradient steps on its support set.

For each task \(T_i\), the inner-loop adaptation updates the model parameters as:
\begin{equation}
    \theta'_i = \theta - \alpha \nabla_{\theta} \mathcal{L}_{T_i}^{\text{support}}(\theta),
\end{equation}
where \(\alpha\) is the inner-loop learning rate, and \(\mathcal{L}_{T_i}^{\text{support}}(\theta)\) denotes the reconstruction loss computed on \(D_i^{\text{support}}\).

The outer-loop meta-optimization minimizes the query losses after adaptation across sampled tasks:
\begin{equation}
    \min_{\theta} \sum_{T_i \sim p(T)} \mathcal{L}_{T_i}^{\text{query}}(\theta'_i),
\end{equation}
where \(\mathcal{L}_{T_i}^{\text{query}}(\theta'_i)\) denotes the loss on \(D_i^{\text{query}}\) evaluated with the adapted parameters \(\theta'_i\).

\subsubsection{Meta-Learning-Based PSI Compression Algorithm}

The meta-training procedure is summarized in Algorithm~\ref{alg:maml}.

\begin{algorithm}[H]
\caption{Meta-Learning Based PSI Compression }
\label{alg:maml}
\begin{algorithmic}
\Require Task distribution \(p(T)\), inner-loop learning rate \(\alpha\), outer-loop learning rate \(\beta\);
\Ensure Meta-learned model parameters \(\theta\);
\State Initialize model parameters \(\theta\);
\While{not converged}
    \State Sample a batch of tasks \(\{T_1, T_2, \dots, T_N\} \sim p(T)\);
    \For{each task \(T_i\)}
        \State Sample support set \(D_i^{\text{support}}\) and query set \(D_i^{\text{query}}\);
        \State Compute adapted parameters: 
        \[
        \theta'_i = \theta - \alpha \nabla_{\theta} \mathcal{L}_{T_i}^{\text{support}}(\theta);
        \]
        \State Evaluate query loss \(\mathcal{L}_{T_i}^{\text{query}}(\theta'_i)\);
    \EndFor
    \State Update meta-parameters:
    \[
    \theta \leftarrow \theta - \beta \nabla_{\theta} \sum_{i=1}^N \mathcal{L}_{T_i}^{\text{query}}(\theta'_i);
    \]
\EndWhile
\State \Return \(\theta\)
\end{algorithmic}
\end{algorithm}

\section{Simulation Results}
\subsection{Training Settings and Procedure}

The meta-training of MCRNet is conducted under fixed IRS settings, with PSI matrices generated from a \(32 \times 32\) IRS array quantized to 4-bit resolution. The PSI matrices are transmitted through a standard independent Rayleigh fading channel model. Each meta-task \(\mathcal{T}_i\) corresponds to a small subset of PSI matrices sampled under the same condition. For each meta-task, the support set contains 100 samples and the query set contains 64 samples, randomly sampled from the available PSI matrices. The loss function used during meta-training is the mean squared error (MSE) between the reconstructed PSI matrix \(\hat{\mathbf{X}}\) and the ground-truth \(\mathbf{X}_{\text{ideal}}\), defined as
\begin{equation}
\mathcal{L}(\theta) = \frac{1}{|\mathcal{D}|} \sum_{(\mathbf{X}, \mathbf{X}_{\text{ideal}}) \in \mathcal{D}} \left\| f_\theta(\mathbf{X}) - \mathbf{X}_{\text{ideal}} \right\|_2^2,
\end{equation}
where \(\mathcal{D}\) denotes the corresponding support or query set.

The model is optimized using the Adam optimizer, with an inner-loop learning rate \(\alpha = 10^{-3}\) and an outer-loop learning rate \(\beta = 5 \times 10^{-4}\). Each inner-loop adaptation consists of one gradient descent step based on the support set, while the outer-loop meta-update aggregates query losses across a mini-batch of 8 meta-tasks. Meta-training is conducted for 1000 iterations with early stopping based on validation loss. All experiments are conducted on a workstation equipped with an NVIDIA RTX A40 GPU.

\subsection{Performance Evaluation}

In this section, we comprehensively evaluate the reconstruction performance, computational efficiency, and few-shot adaptability of the proposed MCRNet framework, compared against representative lightweight baselines including S-GAPSCN and PSCDN.

\subsubsection{Reconstruction and Efficiency Evaluation}

\begin{figure*}[t]
    \centering
    \subfigure[CR = 1/2]{
        \includegraphics[width=0.35\textwidth]{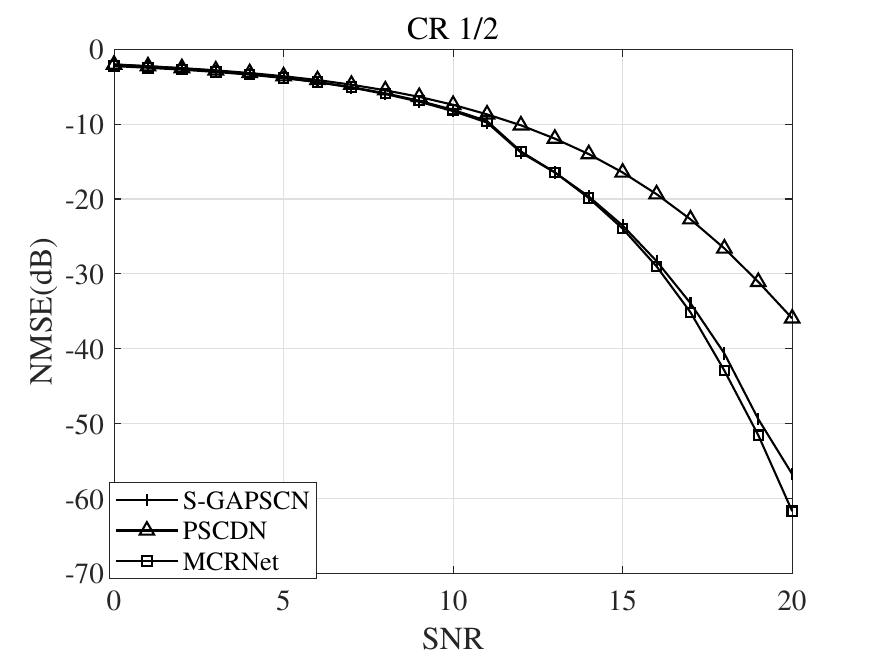}
    }
    \hspace{-20mm}
    \hfill
    \subfigure[CR = 1/4]{
        \includegraphics[width=0.35\textwidth]{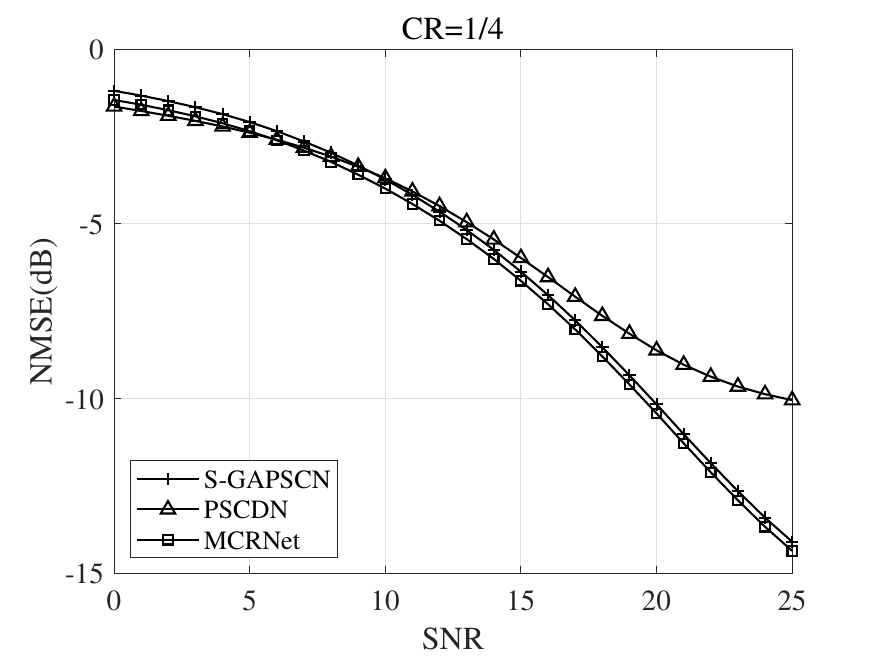}
    }
    \hspace{-20mm}
    \hfill
    \subfigure[CR = 1/8]{
        \includegraphics[width=0.35\textwidth]{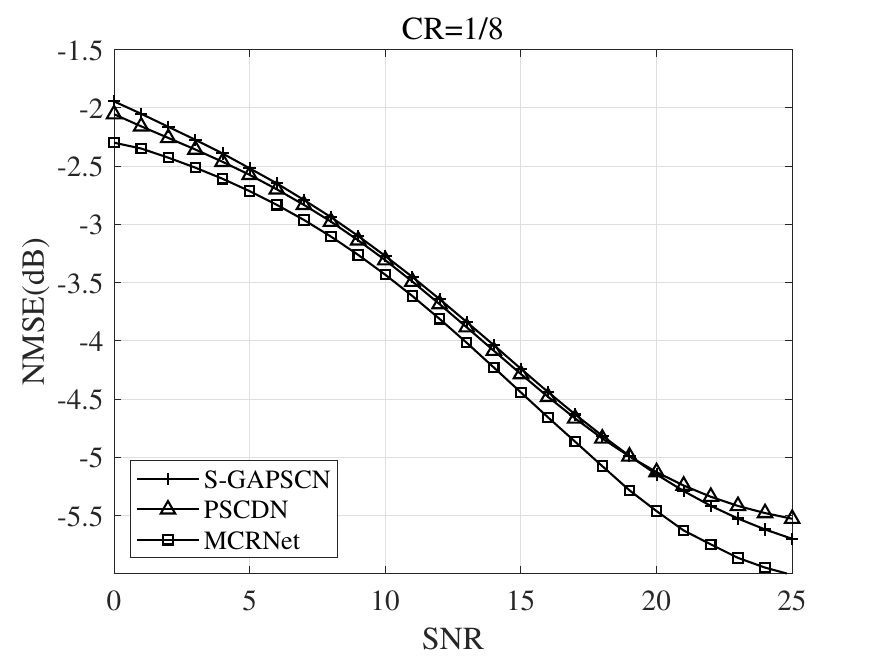}
    }
    \caption{NMSE performance under different compression ratios.}
    \label{cr}
\end{figure*}

The performance of MCRNet is evaluated from two perspectives: reconstruction accuracy and computational efficiency, compared to representative lightweight state-of-the-art (SOTA) baselines, including the S-GAPSCN\cite{XYu2} and the PSCDN\cite{XYu1}. The reconstruction accuracy is measured using the normalized mean square error (NMSE), defined as
\begin{equation}
    \text{NMSE} = \frac{\mathbb{E}\left[\|\hat{\mathbf{X}} - \mathbf{X}_{\text{ideal}}\|_2^2\right]}{\mathbb{E}\left[\|\mathbf{X}_{\text{ideal}}\|_2^2\right]}.
\end{equation}
Fig.~\ref{cr} illustrate the NMSE performance versus SNR under compression ratios of \(1/2\), \(1/4\) and \(1/8\), respectively.

Under moderate compression (CR = 1/4), MCRNet achieves slightly better NMSE performance compared to SGAPSCN and substantially outperforms PSCDN across all SNR regimes. As the compression becomes more aggressive (CR = 1/8), MCRNet maintains competitive reconstruction accuracy while outperforming both baselines in low-to-moderate SNR conditions.

Although the NMSE improvement over existing lightweight baselines is moderate, it is important to note that MCRNet achieves these results with significantly reduced decoder parameter size and inference latency. As summarized in Table~\ref{tab:complexity}, MCRNet reduces the decoder parameters by 44.3\% compared to SGAPSCN and by 74.2\% compared to PSCDN, while accelerating inference by 50.7\% and 71.1\%, respectively. These results clearly demonstrate the practical efficiency advantage achieved by MCRNet. The substantial reduction in model size and inference latency is primarily attributed to two key design principles.

% \begin{figure}[t]
%     \centering
%     \includegraphics[width=0.7\linewidth]{cr14.eps}
%     \caption{NMSE performance comparison under compression ratio CR = 1/4.}
%     \label{fig:cr14}
% \end{figure}

% \begin{figure}[t]
%     \centering
%     \includegraphics[width=0.7\linewidth]{cr18.eps}
%     \caption{NMSE performance comparison under compression ratio CR = 1/8.}
%     \label{fig:cr18}
% \end{figure}

First, the proposed asymmetric encoder-decoder architecture shifts the major computational burden to the encoder side deployed at the BS, which has abundant computational resources, while maintaining a lightweight decoder at the IRS controller by avoiding complex feature transformations during reconstruction. Second, the integration of DWCG modules enables adaptive local feature modulation without relying on costly global attention operations. Specifically, DWCG enables adaptive local feature modulation using lightweight depthwise convolutions and pointwise multiplications, selectively adjusting feature importance at a fine-grained spatial level. This design significantly reduces parameter count and computational overhead compared to conventional global modeling methods such as attention mechanisms.

By combining encoder-side heavy compression with decoder-side efficient modulation, MCRNet achieves a favorable trade-off between reconstruction accuracy and computational efficiency. This enables real-time PSI reconstruction under stringent hardware and latency constraints, making MCRNet highly practical for dynamic and resource-limited IRS-assisted wireless systems.
\begin{table}[H]
\centering
\caption{Decoder Complexity and Inference Time Comparison (CR 1/8)}
\label{tab:complexity}
\begin{tabular}{lcc}
\toprule
\textbf{Method} & \textbf{Decoder Parameters} & \textbf{Inference Time (ms)} \\
\midrule
MCRNet (proposed) & 50,753 & 0.37 \\
SGAPSCN & 91,157 & 0.75 \\
PSCDN & 197,121 & 1.28 \\
\bottomrule
\end{tabular}
\end{table}

\subsubsection{Few-Shot Adaptation Analysis}

In addition, the few-shot adaptation capability of MCRNet is evaluated during the meta-training process.

Benefiting from the MAML framework, MCRNet is capable of rapidly adapting to unseen IRS configurations using only 100 support samples per task. In contrast, conventional methods such as SGAPSCN and PSCDN typically require over 1000 samples to achieve effective task-specific training.

This highlights the strong generalization capability enabled by the proposed meta-learned initialization, which significantly reduces retraining overhead while maintaining high reconstruction performance. Such few-shot adaptability is critical for dynamic wireless environments, where IRS conditions may vary frequently, and efficient model updating is required without relying on large-scale retraining.

\section{Conclusion}

This paper has proposed MCRNet, a meta-learning-enhanced lightweight autoencoder framework for PSI compression in IRS-assisted wireless systems. By combining an asymmetric encoder-decoder design with adaptive local feature modulation via DWCG and few-shot adaptation through MAML, MCRNet has achieved competitive reconstruction accuracy while significantly reducing model size and inference latency. These results have demonstrated the practicality and scalability of MCRNet for real-time and resource-constrained IRS-assisted deployments.

\end{document}